\documentclass[%
 reprint,
showpacs,
 amsmath,amssymb,
 aps,
sort&compress,
pra,
]{revtex4-1}

\usepackage{natbib}

\usepackage{graphicx,subfigure}
\usepackage{dcolumn}
\usepackage{bm}

\usepackage{xparse}
\usepackage{tikz}
\usetikzlibrary{matrix,backgrounds}
\pgfdeclarelayer{myback}
\pgfsetlayers{myback,background,main}

\tikzset{mycolor/.style = {line width=1bp,color=#1}}%
\tikzset{myfillcolor/.style = {draw,fill=#1}}%

\NewDocumentCommand{\highlight}{O{blue!40} m m}{%
\draw[mycolor=#1,#1] (#2.north west)rectangle (#3.south east);
}

\NewDocumentCommand{\fhighlight}{O{blue!20} m m}{%
\draw[myfillcolor=#1,#1] (#2.north west)rectangle (#3.south east);
}

\begin{document}
\newcommand\hlight[1]{\tikz[overlay, remember picture,baseline=-\the\dimexpr\fontdimen22\textfont2\relax]\node[rectangle,fill=blue!50,rounded corners,fill opacity = 0.2,draw,thick,text opacity =1] {$#1$};}


\title{Exploring the limits of multiplexed photon-pair sources for the preparation of pure single-photon states}

\author{Robert J.A. Francis-Jones}
 \email{r.j.a.francis-jones@bath.ac.uk}
\author{Peter J. Mosley}%
\affiliation{Centre for Photonics and Photonic Materials, Department of Physics, University of Bath, Bath, BA2 7AY}%




\date{\today}

\begin{abstract}
Current sources of heralded single photons based on nonlinear optics operate in a probabilistic manner. In order to build quantum-enhanced devices based around the use of single photons, compact, turn-key and deterministic sources are required. A possible solution is to multiplex a number of sources to increase the single-photon generation probability and in so doing reducing the waiting time to deliver large numbers of photons simultaneously, from independent sources. Previously  it has been shown that, in the ideal case, 17 multiplexed sources allow deterministic generation of heralded single photons [Christ and Silberhorn, \textit{Phys. Rev. A} \textbf{85}, 023829 (2012)]. Here we extend this analysis to include undesirable effects of detector inefficiency and photon loss on a number of multiplexed sources using a variety of different detectors for heralding. We compare these systems for fixed signal-to-noise ratio to allow a direct comparison of performance for real-world heralded single photon sources.

\begin{description}
\item[PACS numbers]42.65.Lm, 42.50.Ar, 42.50.Dv, 42.50.Ex
\end{description}
\end{abstract}

\maketitle


\section{\label{sec:Introduction}Introduction}

By exploiting the principles of superposition and entanglement, quantum-enhanced technologies such as quantum computing, quantum cryptography and quantum enhanced measurements can be realised~\cite{OBrien2009Photon_Quantum_Technologies}. Photonics is at the forefront of such research, using single photons as quantum bits. However preparing single photon states is not a trivial task: sources must produce one and only one photon on demand, and every photon must be in an identical pure quantum state \cite{Knill2001A-scheme-for-efficient-quantum, Varnava2008How-Good-Must-Single}

The most widely used single photon sources are based around nonlinear frequency conversion. These sources utilise the nonlinear optical response of a medium to an intense laser pulse to generate pairs of daughter photons termed the signal and idler. The two main mechanisms are parametric downconversion (PDC) in bulk crystals~\cite{Hong1986Experimental_Realisation} and four-wave mixing (FWM) in photonic crystal fibre (PCF)~\cite{Fiorentino2002All_Fibre_Source, Alibart2006Photon_Pair_Generation, Clark2011-Intrinsically-narrowband-pair}. Despite their ubiquity, PDC and FWM sources suffer from some limitations. In the majority of sources, photons are generated in many correlated spectral and spatial modes resulting in heralded single photons in mixed quantum states. To avoid the use of lossy filters, care must be taken to engineer the generating medium to remove these correlations \cite{Grice2001-Eliminating-frequency-and}; this area has seen considerable progress in recent years~\cite{Mosley2008-Heralded-Generation-of-Ultrafast, Garay-Palmett2007-Photo-pair-state-preparation, Cohen2009-Tailored-Photon-Pair-Generation, Soller2010-Bridging-visible-and-telecom, Soller2011High_Performance}. Nevertheless, even when operating in this regime, the fundamental mechanism of photon-pair generation is spontaneous; this enforces a severe limit on the probability of generating a pair on any single pump pulse (typically around 1\%). Hence source performance is far from the deterministic ideal.

Active multiplexing of several photon-pair sources provides an attractive route to bypassing the problem of non-deterministic performance while retaining many of the advantages of robust, room-temperature photon-pair sources. By connecting a number of sources to a network of optical switches, a photon from any one of the sources can be routed to a common output. Spontaneous photon-pair sources are particularly amenable to this technique as the result of a heralding detection of one photon can be fed forward to set the switches to route the other photon to the output. A number of multiplexed systems have been proposed both in the spatial domain (using separate spatial modes or sources) \cite{Migdall2002Tailoring-single-photon-and-multiphoton, Shapiro2007On_Demand_Single_Photon} and temporal domain (pulse recycling) \cite{Jeffrey2004Towards-a-periodic-deterministic, Mower2011Efficient-generation-of-single, Glebov2013Deterministic-generation-of-single, Schmiegelow2013Multiplexin_photons_with}; more advanced concepts include implementation in the spectral degree of freedom \cite{Kumar2013Spectrally-multiplexed-and-tunable-wavelength} or multiplexing by storing photons in quantum memories \cite{Nunn2013Enhancing-Multiphoton-Rates}. Experimental demonstrations have been carried out of passive (non-switched) temporal multiplexing through sub-division of pump laser pulses \cite{Broome2011Reducing-multi-photon-rates}, active noise reduction through the use of a fast switch \cite{Brida2011Experimental-realization-of-a-low-noise}, and two- and four-source spatially-multiplexed sources \cite{Ma2011Experimental_Generation_of, Collins2013Integrated-spatial-multiplexing}. Although experimental implementations to date have been limited, it has been shown theoretically that by multiplexing several non-deterministic sources together deterministic operation can be approached~\cite{Christ2012Limits_on_the, Glebov2013Deterministic-generation-of-single}.

Multiplexed sources are limited by the inefficiency and speed of both detectors and switches. Current silicon single-photon avalanche photodiodes have peak efficiencies of 70\% and are binary, providing no information about the number of photons that hit the detector. Photon-number resolving (PNR) detectors with high efficiencies over the range of $800-850$nm are available by utilising transition edge sensors (TES) though their timing jitter is large \cite{Fukuda2011Titanium-Based_Transition_Edge_Sensor, Lita2008Counting_near_infrared,Lita2010Superconducting_transition_edge, Rosenberg2005Noise_Free_High_Efficiency}. Current $2$-to-$1$ switches operating around $1550$nm have efficiencies in the range of $70-80\%$ with repetition rates around $1$MHz. The detector response time and switch bandwidth limit the laser repetition rate and hence maximum rate at which photon pairs can be detected and routed through the switch network.

In this paper we consider theoretically systems of photon-pair sources multiplexed in the spatial domain and analyse their performance in the presence of switch loss and detetor inefficiency. We investigate the relative trade-off between heralding rate and the quality of the state generated, allowing a comparison to the results contained in Christ et al  \cite{Christ2012Limits_on_the}. We initially consider three cases of single sources utilising binary, PNR, and pseudo-PNR detectors. From this basis we then study the spatial multiplexing of a number of these sources in an integrated network to investigate how closely we may approach deterministic operation. By including the effects of detector inefficiency and different detector types, a meaningful comparison of different systems becomes non-trivial. We must fix one indicator of source performance: here we choose the signal-to-noise ratio (SNR) -- the fraction of single-photon to multi-photon states delivered at the source output. Fixing the SNR requires us to use numerical methods to compare source performance. Although we explicitly consider spatially-multiplexed sources, the same analysis can be straightforwardly extended to multiplexing in the temporal domain.

\section{\label{sec:PhotonPairs} Photon-Pair Generation}
Figure~\ref{fig:SingleBinSource} shows the pair generation process for a general photon-pair source based on either PDC or FWM. The output state can be written as a superposition of photon number:
\begin{equation}
	\label{eq:StateVector}
		|\Psi\rangle = a_{0}|0_{s},0_{i}\rangle + a_{1}|1_{s},1_{i}\rangle + a_{2}|2_{s},2_{i}\rangle + \cdots.
\end{equation}
Here we assume that the source has been engineered to emit signal and idler photons into only two spatio-temporal modes and the arms are correlated only in photon number. As a result the probability amplitudes are described by thermal statistics~\cite{Loudon1973The_Quantum_Theory}:
\begin{equation}
	\label{eq:Thermal}
	|a_{n}|^2 = p_{th}(n) = \frac{1}{(\bar{n} + 1)}\left(\frac{\bar{n}}{\bar{n} + 1}\right)^n,
\end{equation}
where $\bar{n}$ is the mean photon number per pump pulse. The statistical distribution of photon pairs is plotted in Figure~\ref{fig:SingleBinSource}; note that for a thermal source the probability per pulse of generating a single pair cannot exceed 0.25.

The dominant vacuum component $|0_{s},0_{i}\rangle$ is typically removed through the detection of one photon in each pair to herald the presence of its twin. The remaining photon is projected into a sum of photon number states with weightings given by the set of probability amplitudes $\{a_{n}\}$. The overall quality of the heralded state is limited both by the contributions of higher-order photon-number terms and the residual vacuum component resulting from loss of the photon. We define a signal-to-noise ratio (SNR) as the relative contribution of single-photon to multi-photon terms in the heralded state:
\begin{equation}
	\label{eq:SNR}
		SNR = \frac{P(1)}{\sum\limits_{n=2}^{\infty}P{(n)}}.
\end{equation}
where $P(n)$ is the probability per pulse of delivering $n$ photons in the output arm. In order to compare meaningfully different source configurations we set the SNR equal for each. As a metric for overall source performance, we use the fidelity of the delivered state with a pure single-photon Fock state \cite{Christ2012Limits_on_the}.

\section{\label{sec:IndividualSources}Individual Sources}
First we consider an individual single photon source utilising 3 different detector types: binary, photon number resolving (PNR) and pseudo-PNR, with an output state vector defined as in Eq.~\eqref{eq:StateVector} and illustrated in Fig.~\ref{fig:SingleBinSource}. These single sources will then be used as building blocks from which a multiplexed system can be constructed. The heralding measurement performed by the detector placed in the signal arm is described by a set of positive operator valued measure (POVM) elements.

\begin{figure}
			\subfigure[]{
				\label{fig:SingleBinSource}
				\includegraphics[width=0.45\textwidth]{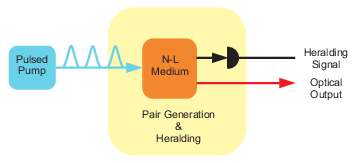}
				}
			
			\subfigure[]{
				\label{fig:BarPlot}
				\includegraphics[width=0.22\textwidth]{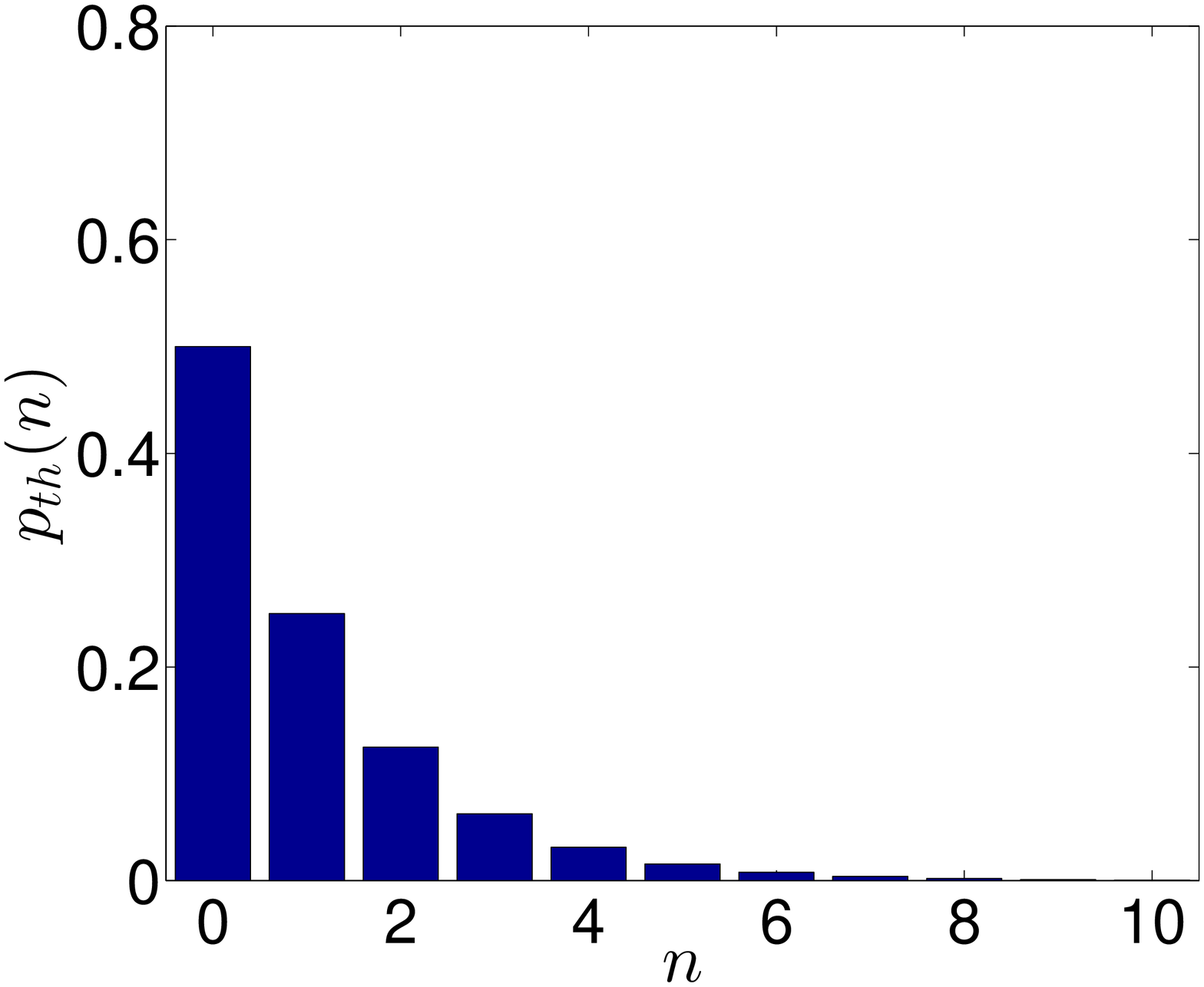}
				}
			\hspace*{\fill}
			\subfigure[]{
				\label{fig:ThermalStats}
				\includegraphics[width=0.22\textwidth]{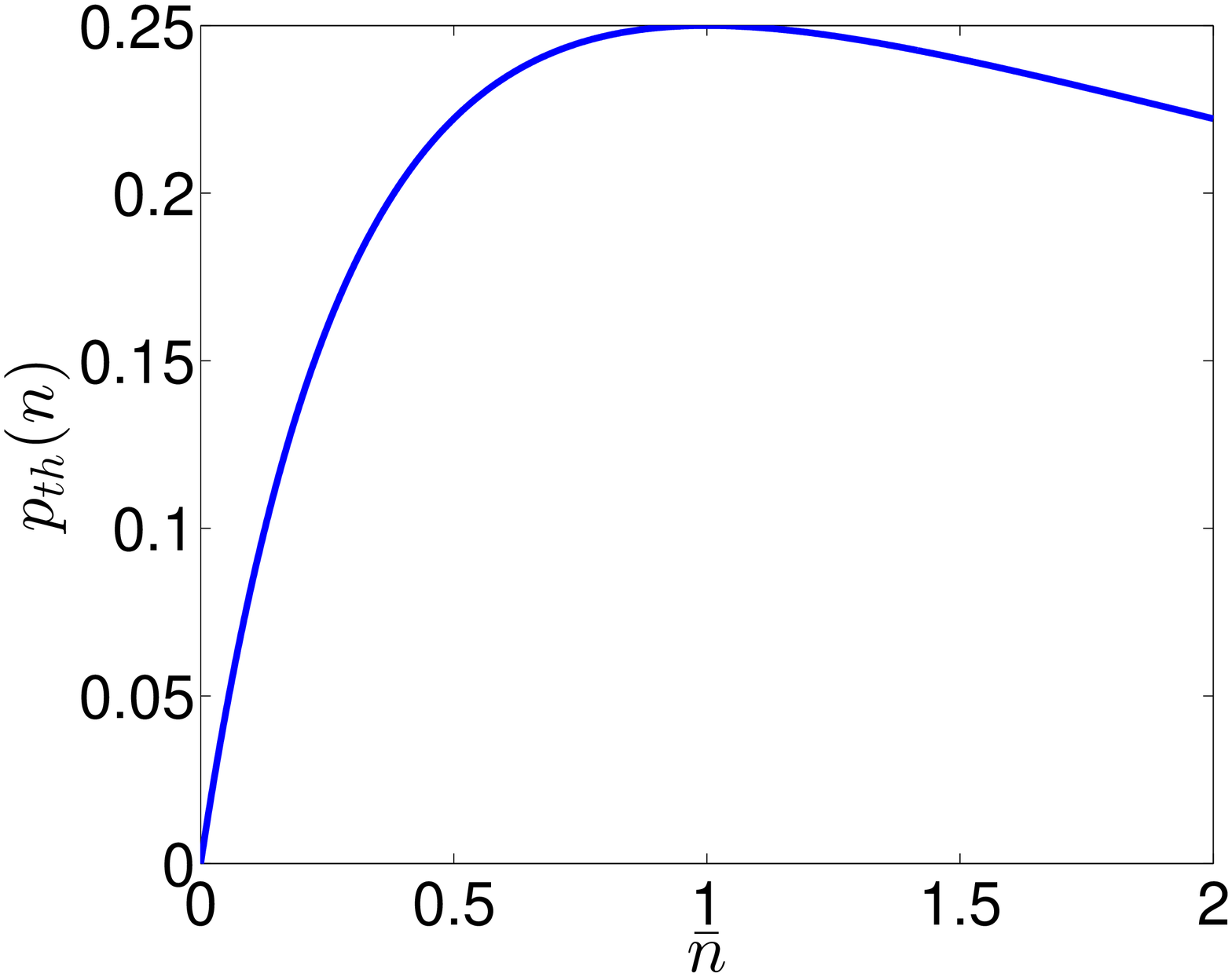}
				}
\caption{a.) A single photon source modelled as a black box. A pulsed laser system pumps a non-linear medium (orange). The resulting photon pair is split into the signal (black arrow) and idler (red arrow). A detector is placed in the signal arm to herald the presence of photons in the idler arm. b.) Photon number statistics of a thermal distribution with $\bar{n} = 1$. c.) Single photon pair probability with increasing $\bar{n}$.}
\label{fig:SingleBinSource}
\end{figure}

The general form of such a POVM operator for the outcome `$n$'  can be written as 
\begin{equation}
	\hat{\Pi}(n) = \sum\limits_{N=n}^{\infty}p_{det}(n|N)|N\rangle\langle N |
	\label{POVM}
\end{equation}
where $p_{det}(m|N)$ is the conditional probability of the detector recording result `$n$' as the result of $N$ photons at the input. This conditional detection probability is dependent on the detector type and mechanism of detection. 

For a binary detector where the only two possible results are ``click" and ``no click" the POVM operator set contains only the two elements arising from the two possible results of detection:
\begin{equation}
	p_{det}(\text{``click''}|N) = [1-(1-\eta_{d})^N],
\end{equation} 

\begin{equation}
	p_{det}(\text{``no-click"}|N) = (1-\eta_{d})^N,
\end{equation}
where $\eta_{d}$ is the efficiency of the detector. Note that this is a lumped efficiency that accounts for all loss in the channel leading to the detector.

PNR detection can be approximated by splitting the detector input into a number of individual spatial or temporal modes and monitoring each mode with a binary detector ~\cite{Achilles2003Fibre_assisted_detectio_with, Rohde2005Non_determinsti_approximation_of_photon, Coldenstrodt-Ronge2010A_proposed_testbed, Worsley2009Absolute_efficiency_estimation}. Typically the splitting is performed by a network of beam splitters or 50:50 fibre couplers and the probability of a photon occupying a mode is proportional to the inverse of the number of modes for equal splitting probabilities. The limitation of such a pseudo-PNR detector is that its operation is non-deterministic: more than one photon can end up in the same detector mode. For a detector with some photon number resolving capability the set of POVM elements extends over all possible values of photon number n but limited by the number of detection modes present. A pseudo-PNR detector constructed as in~\cite{OSullivan2008Conditional_Preparation} whereby the input state is split across a number of detecton modes $M$, each monitored by a standard binary detector, can be described by a set of POVM elements for measuring n photons with a conditional detection probabilities given by
\begin{equation}
	p_{det}(n|N) = {M \choose n}\sum\limits_{j=0}^{n} (-1)^j  {n \choose j} \left( (1-\eta_{d}) + \frac{\eta_{d}(n - j)}{M}\right)^{N}.
\end{equation}
By allowing the number of modes to tend to infinity the POVM elements of a ideal photon-number resolving detector with detection efficiency $\eta_{d}$ are recovered yielding conditional detection probabilities of the form \cite{OSullivan2008Conditional_Preparation},
\begin{equation}
	p_{det}(n|N) = {N \choose n} \eta_{d}^n (1-\eta_{d})^{N-n}.
\end{equation}

The resultant reduced density matrix of the heralded idler state following the detection of $n_{s}$ photons is then found by projecting the relevant POVM element onto the single-mode pair state and tracing over the detected signal photon:
\begin{equation}
	\hat{\rho}_{i}(n_{s}) = \frac{tr_{s}(\hat{\Pi}_{det.}|\psi\rangle\langle\psi|)}{\langle\psi|\hat{\Pi}_{det.}|\psi\rangle}.
\end{equation}
This then defines an ensemble of possible reduced density matrices of the heralded idler state given the detection of $n_{s}$ photons in the signal arm. The signal-to-noise of the heralded state can be determined by selecting the reduced density matrix corresponding to a successful detection event and calculating the relative contribution of the single-photon and higher-order terms. In the case of a PNR detector this becomes:
\begin{equation}
	SNR = \frac{\langle{1_{i}}|\hat{\rho}_{i}(n_{s} = 1)|1_{i}\rangle}{\sum\limits_{k=2}^{\infty}\langle k_{i}|\hat{\rho}_{i}(n_{s} = 1)|k_{i}\rangle},
\end{equation}
whereas for binary detection it becomes:
\begin{equation}
	SNR = \frac{\langle{1_{i}}|\hat{\rho}_{i}(\text{``click"})|1_{i}\rangle}{\sum\limits_{k=2}^{\infty}\langle k_{i}|\hat{\rho}_{i}(\text{``click"})|k_{i}\rangle},
\end{equation}
as the higher-order componenets cannot be distinguished by the heralding detection. The SNR as a function of pair generation probability for a single pair generation source for each of the three detectors is displayed in Fig.~\ref{fig:16SourceSNR}. We then set the SNR to a fixed value to allow us to make direct comparisons between systems utilising different detectors.

The fidelity of the heralded idler state can be found by determining the overlap of the reduced density matrix after heralding with an ideal single photon Fock state:
\begin{equation}
	F = \frac{\langle 1|\hat{\rho}_{i}|1\rangle}{\text{Tr}\{ \hat{\rho}_{i}\}}.
\end{equation}
We identify a successful outcome of the source as a single heralding (PNR) or ``click" event (binary) leading to a single heralded photon. To find the overall probability per pulse of successfully producing a heralded single photon from the source, $p(\text{success})$, we first determine the probability of a succesful heralding detection, $p(\text{heralding})$ and then multiply by the fidelity of the heralded idler state. For a PNR detector $p(\text{heralding}) = p(n_{s} = 1)$ and we have:
\begin{equation}
	\label{eq:PNR_pSuccess}
	p(\text{success}) = p(n_{s} = 1)\frac{\langle 1|\hat{\rho}_{i}(n_{s} = 1)|1\rangle}{\text{Tr}\{ \hat{\rho}_{i}(n_{s} = 1)\}},
\end{equation}
whereas for a binary detector $p(\text{heralding}) = p(\text{``click"})$ and:
\begin{equation}
	\label{eq:Binary_pSuccess}
	p(\text{success}) = p(\text{``click"})\frac{\langle 1|\hat{\rho}_{i}(\text{``click"})|1\rangle}{\text{Tr}\{ \hat{\rho}_{i}(\text{``click"})\}}.
\end{equation}
From these success probabilities the mean waiting time, $t_{\text{wait}}$, to deliver $N_{p}$ independent photons from $N_{p}$ independent systems can be found:
\begin{equation}
	t_{\text{wait}} = \left(\frac{p(\text{success})}{R_{p}}\right)^{N_{p}},
\label{eq:wait}
\end{equation}
where $R_{p}$ is the laser repetition rate. We use the waiting time to compare directly the performance of different sources at a constant SNR.

\section{\label{sec:Multiplexing2} Spatial Multiplexing of independent sources}

\subsection{Two independent sources}

\begin{figure}
	\begin{center}
		\begin{subfigure}[]{
				\label{fig:TwoWayMulti}
				\includegraphics[width=0.45\textwidth]{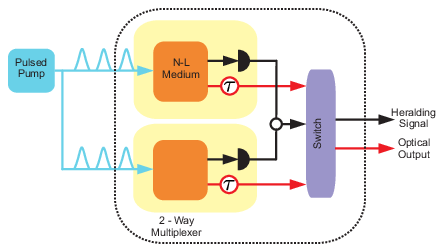}
				}
		\end{subfigure}

		\begin{subfigure}[]{
				\label{fig:NWayMulti}
				\includegraphics[width =0.45\textwidth]{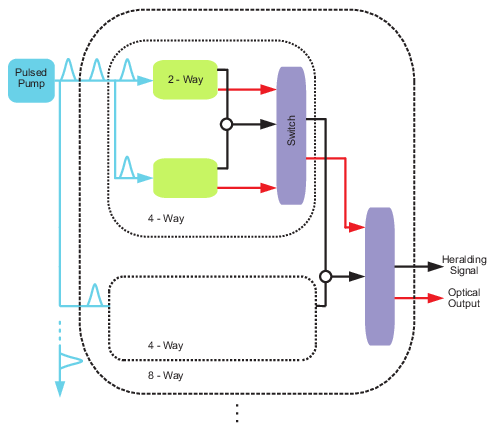}
				}
		\end{subfigure}
	\end{center}
	\caption{Schematic of a multiplexed system. See text for details}
	\label{fig:MultiSource}
\end{figure}

We mulitplex the individual source building blocks defined earlier in a pair wise fashion \cite{Shapiro2007On_Demand_Single_Photon, Ma2011Experimental_Generation_of} as detailed in Fig.~\ref{fig:MultiSource} (a). The outputs of two individual sources each with its own heralding detector are coupled to long lengths of optical fibre to incur a time delay before being routed to a 2-to-1 optical switch. The state of the switch is controlled by the detectors placed in the heralding arms, hence photons in the signal arm must be delayed to allow time for the switch to be set before the heralded photons arrive. We make the assumption that the switch is faster than the repetition rate of the laser source so that the transmission of pairs from each pulse can be selected independently; hence we limit the repetition rate to 1\,MHz. Therefore, when one detector provides a sucessful heralding signal, the switch routes the heralded state of the corresponding source to the output and the channel from the other source to the output remains closed.

For both individual sources we define a heralding probability vector in which each row  corresponds to a particular POVM element. For PNR detection we have:
\begin{equation}
	\vec{p_{H}} = \left( \begin{array}{c}
			 \langle\psi|\hat{\Pi}(0)|\psi\rangle \\
			 \langle\psi|\hat{\Pi}(1)|\psi\rangle \\
			 \langle\psi|\hat{\Pi}(2)|\psi\rangle \\
					\vdots \\
			\langle\psi|\hat{\Pi}(n)|\psi\rangle \\
			\end{array} \right),
\end{equation}
where the first row corresponds to zero photons detected, the second row corresponds to 1 photon detected and so on \cite{Christ2012Limits_on_the}. The equivalent vector for binary detection contains only two elements. From $\vec{p_H}$ we can determine all the possible combinations of detection events across the two sources by taking the Kronecker product of the two heralding probability vectors, yielding an $n \times n$ matrix:
\begin{equation}
		p_{H}^{(2)}  = \vec{p_{H}}^{T} \otimes \vec{p_{H}}
		= \begin{tikzpicture}[baseline=-\the\dimexpr\fontdimen22\textfont2\relax ]
					\matrix (m)[matrix of math nodes,left delimiter=(,right delimiter=)]{
						    {P_{00}} & {P_{10}} & {P_{20}} & {\dots} & {P_{n0}}\\ 
						    {{P_{01}}} & {P_{11}} & {P_{21}} & {\dots} & {P_{n1}}\\
				   		    {P_{02}} & {{P_{12}}} & {P_{22}} & {\dots} & {P_{n2}}\\
						   \vdots & \vdots & \vdots & \ddots & \vdots \\
						    {P_{0n}} & {P_{1n}} & {P_{2n}} & \dots & {P_{nn}}\\ };
					\begin{pgfonlayer}{myback}
					\fhighlight{m-1-2}{m-5-2}
					\fhighlight{m-2-1}{m-2-5}
					\end{pgfonlayer}
					\end{tikzpicture}.
\label{eq:KronProd}
\end{equation}
By summing elements we can recover a heralding probability vector for the two-source multiplexed system. All those elements highlighted in blue correspond to the probability of at least one of the sources singalling a successful heralding event; thereby the switch selects the density matrix $\hat{\rho_{i}}(n_{s} = 1)$ from the ensemble for the combined system. The remaining elements all correspond to no successful heralding events from either source in which case the switch remains closed and the output is in the vacuum state.

When a successful heralding event occurs, we need to take account of the loss in the components routing photons from source to switch and loss through the switch itself. To do this the delay line is defined as having an efficiency $\eta_{\tau}$ and the switch efficiency as $\eta_{s}$. The total loss from the point of generation to the output of the complete multiplexed system can be treated as a unbalanced beam splitter with a transmission coefficient, $T  = \sqrt{\eta}$, and reflectivity coefficient of $R = \sqrt{1-\eta}$ where the total concatenated loss $\eta = \eta_{\tau}*\eta_s^{N_{s}}$ and $N_{s} = \log_{2}(N)$ is the number of switches required for N sources. The transmitted state can be found by applying the beam splitter transformation for an input state containing $n_i$ photons:

\begin{equation}
	|n_{i}^{A},0_{i}^{B}\rangle = \sum\limits_{p=0}^{n} { n \choose p}^{\frac{1}{2}} {\eta^{\frac{p}{2}}}{(1-\eta)^{\left(\frac{n-p}{2}\right)}}|p_{i}^{C},(n-p)_{i}^{D}\rangle,
	\label{eq:BeamSplitterTransform}
\end{equation}
where $A,B$ and $C,D$ label the input and output ports of the beam splitter respectively. The reduced density operator for the idler arm incorporating photon loss is then found by tracing over the loss mode $D$ $|(n-p)_{i}^{D}\rangle$ and then subsequent renormalisation. 

In order to find the overall success probability we calculate the fidelity of density matrix corresponding to a succesful heralding detection event and multiply by the probability per pulse of making a succesful heralding detection. 

\subsection*{\label{sec:MultiplexingN}Extension to N independent sources}

In order to extend this to an arbitrary number of multiplexed sources, we cascade pairs of sources as shown in Fig.~\ref{fig:NWayMulti}. We determine the heralding and success probabilities from four sources and so on using the same method as for two.

All heralding signals corresponding to those detection events which result in either zero or $n>2$ pairs are then ignored as we are only interested in the generation of single photons via a single heralding signal. 

\section{\label{Discussion} Discussion}

\subsection{Signal-to-noise}

\begin{figure} 
		\subfigure[]{
			\label{fig:16SourceSNR}
			\includegraphics[width=0.22\textwidth]{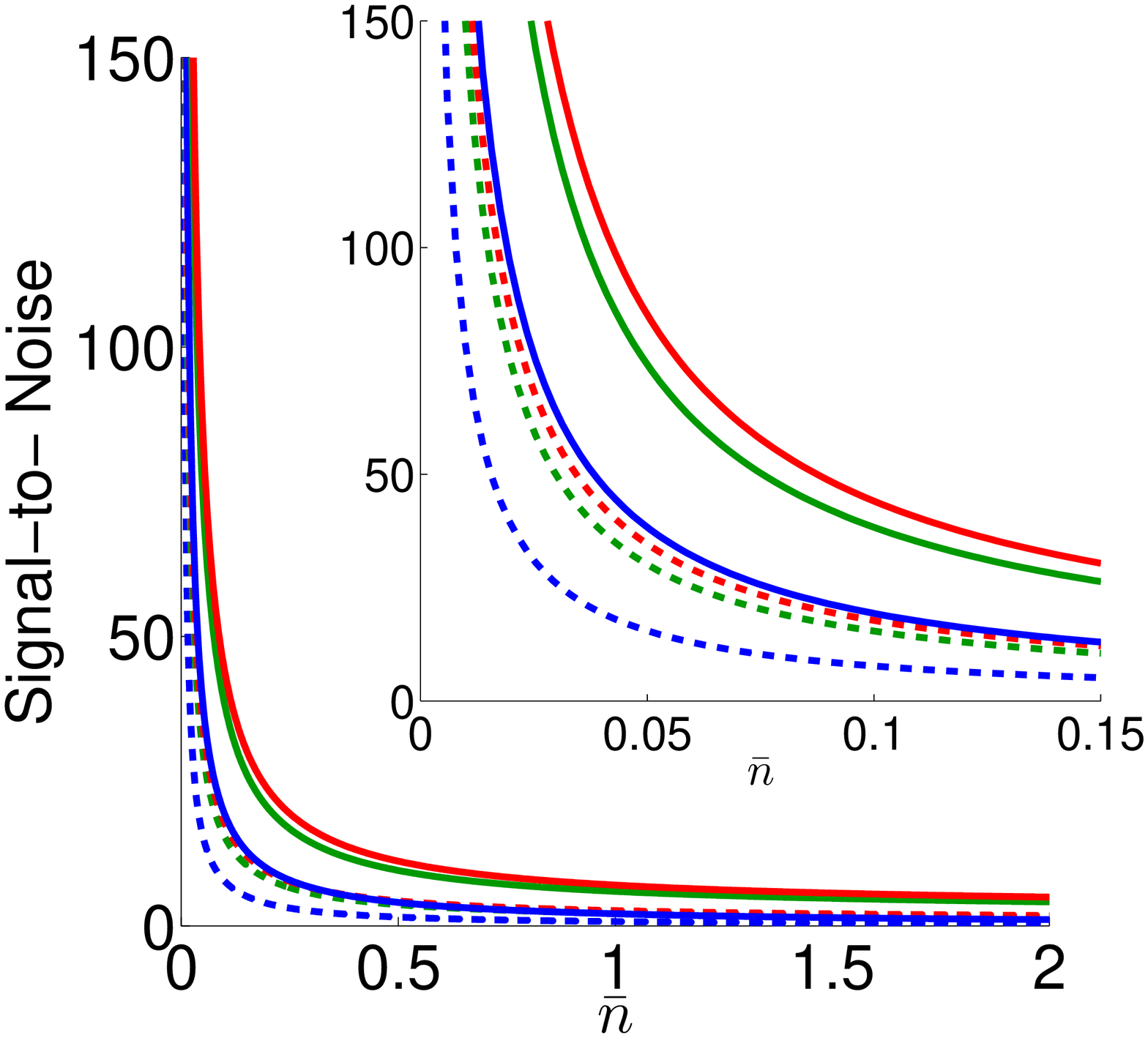}
			}
		\hspace*{\fill}
		\subfigure[]{
			\label{fig:16SourceBar}
			\includegraphics[width=0.22\textwidth]{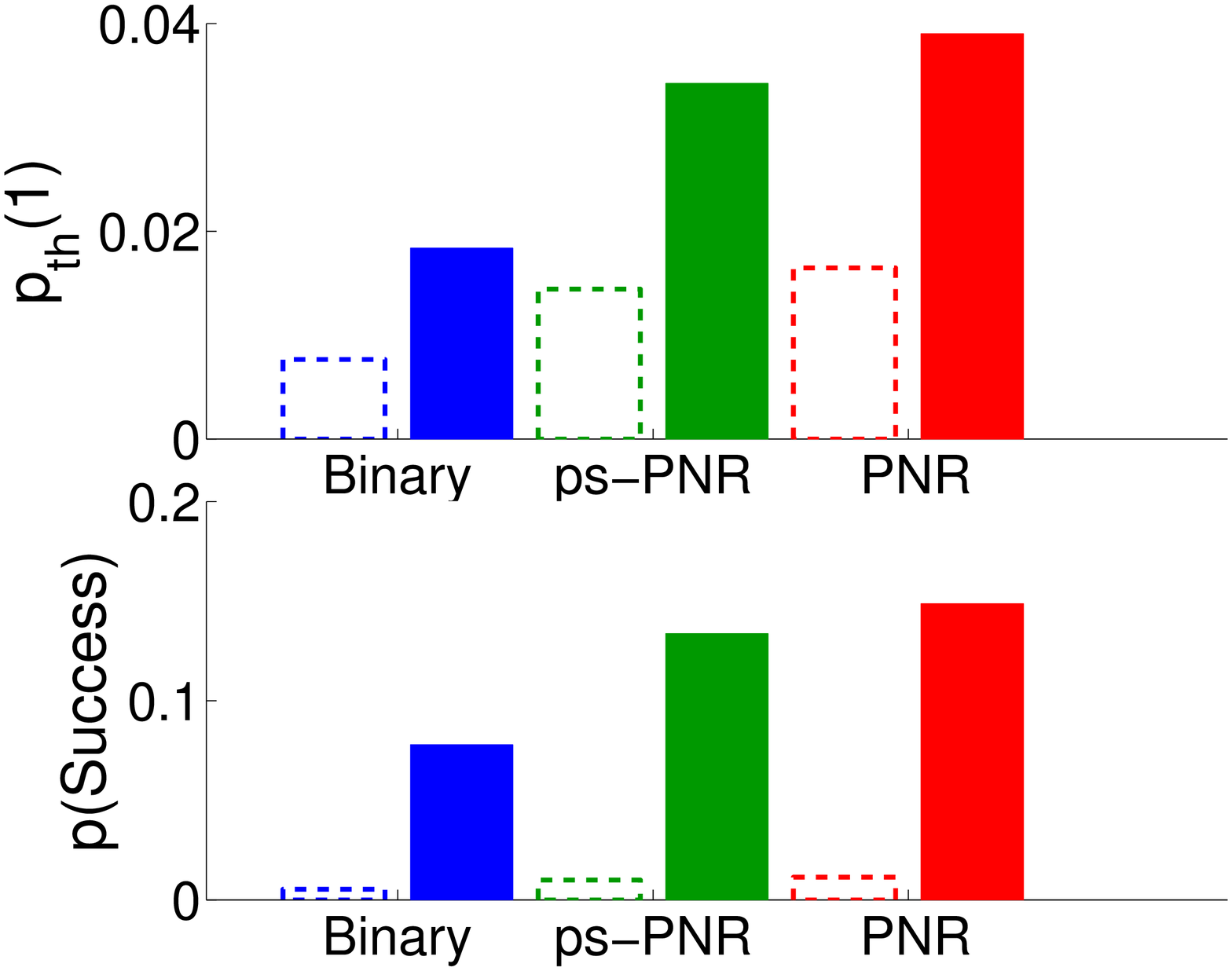}
			}
		
		\subfigure[]{
			\label{fig:16_SourceSNR_nd}
			\includegraphics[width=0.22\textwidth]{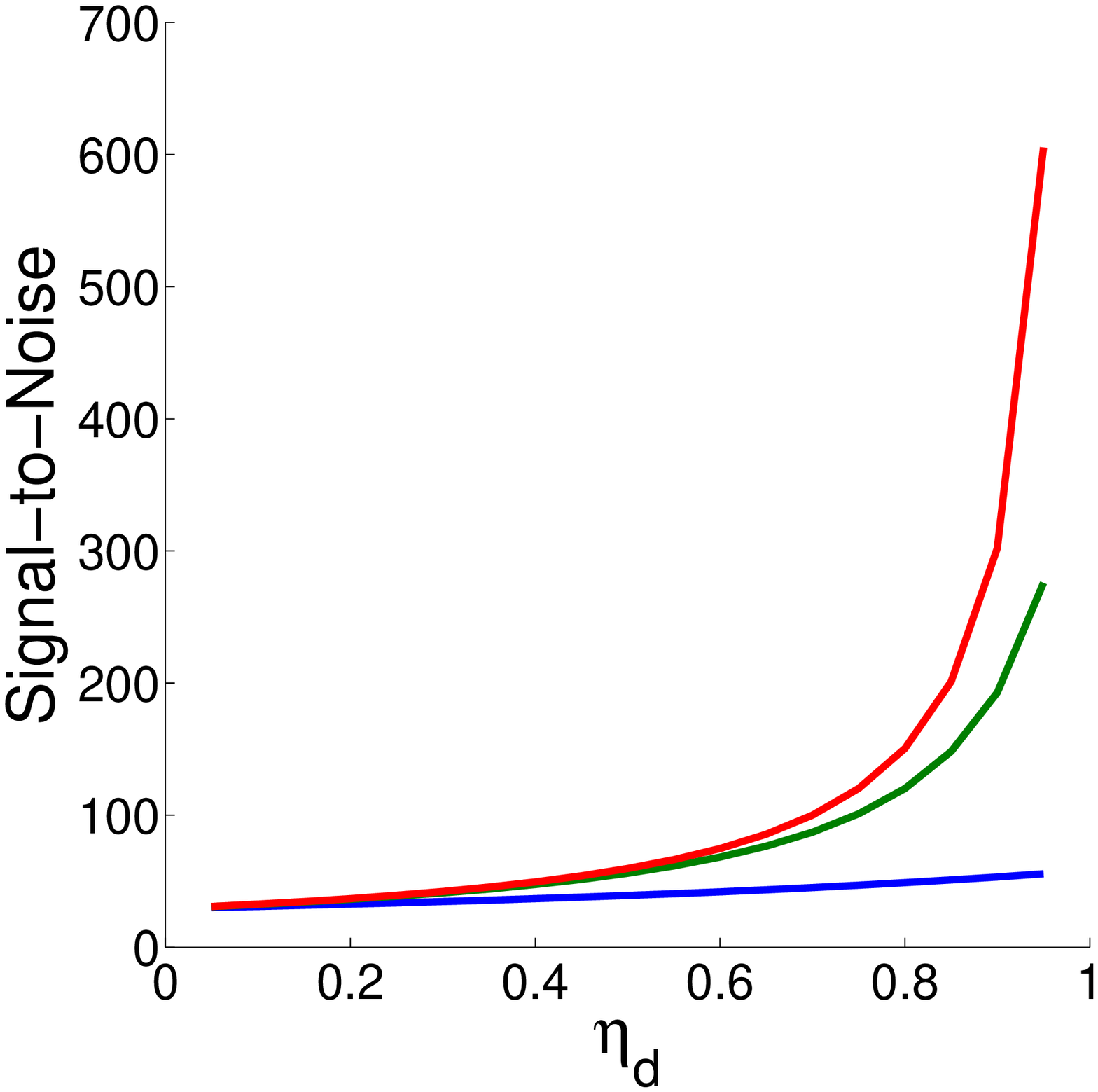}
			}
		\hspace*{\fill}
		\subfigure[]{
			\label{fig:16_SourceSNR_ColorMap}
			\includegraphics[width=0.22\textwidth]{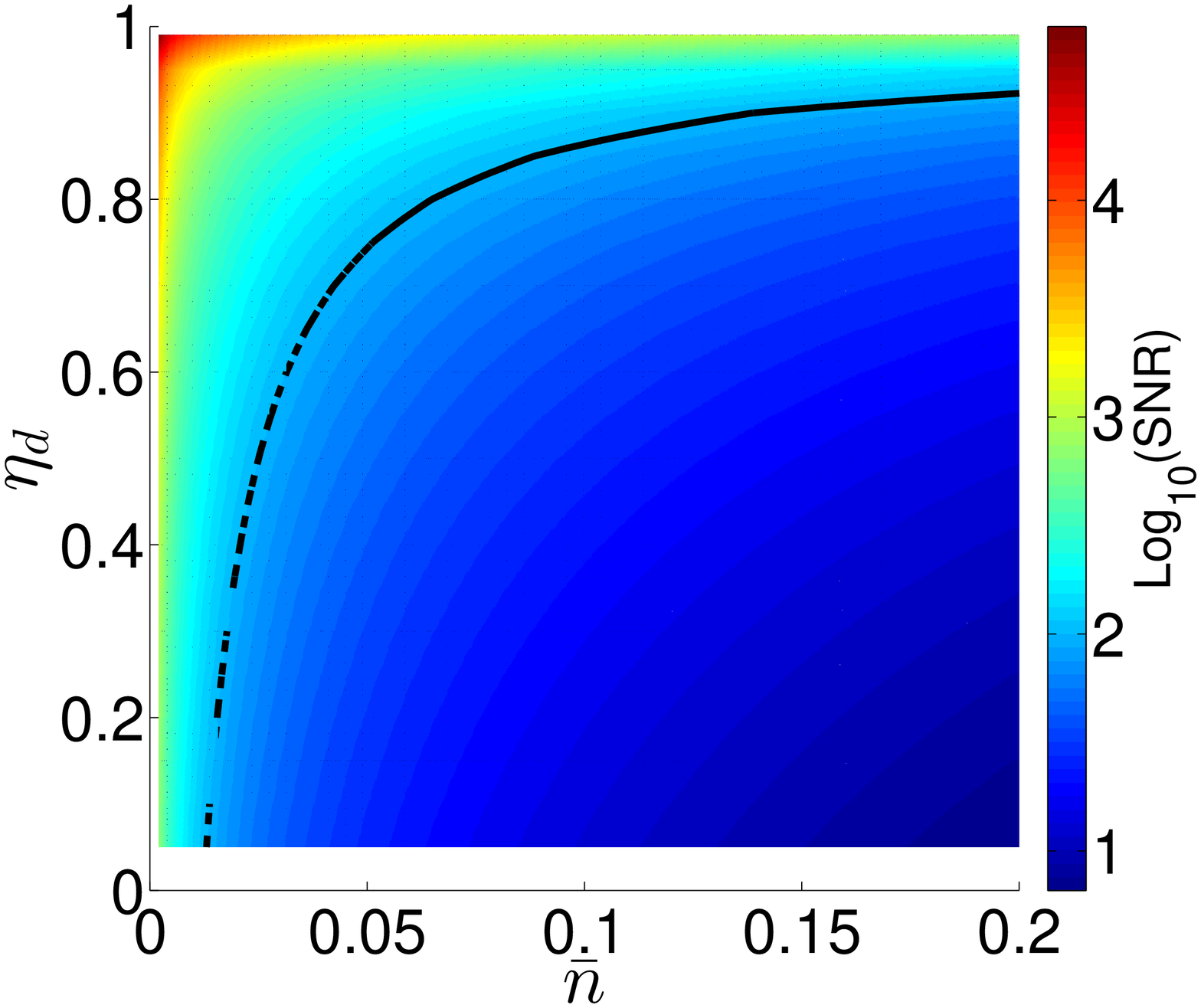}
			}
	\caption{a.) Comparison of achievable SNR for individual (dashed) and 16-source multiplexed systems (solid) for three different heralding detectors: Binary (blue), pseudo-PNR (green) and PNR (red), (inset) performance at low average photon numbers. b.) (top) 1-pair generation probabilities at mean photon numbers that yield a SNR of 100, (bottom) the corresponding probability of heralding and delivering a single photon. c.) Dependence of SNR on detection efficiency for $\bar{n}$ yielding a SNR of 100 at $\eta_{d} = 70\%$. d.) SNR as a function of detection efficiency and average photon number; contour at SNR = 100.}
\end{figure}

Figure~\ref{fig:16SourceSNR} shows the calculated SNR for individual sources and multiplexed systems containing 16 individual sources, using the three different detectors defined in Section~\ref{sec:IndividualSources}. At high mean photon numbers the overall heralded idler state is dominated by large contributions from multi-photon generation resulting in a low SNR. As the mean photon number is reduced the SNR increases as the noise is reduced to a more mangable level. PNR detectors offer a greater SNR for fixed mean photon number due to their ability to discriminate heralding events from multiple photon pairs unlike binary detectors. There is an overall increase in the SNR in moving to a multiplexed system as the overall probability of succesfully of heralding a single photon increases. We compare the different systems and detectors by fixing the SNR at value of 100 and adjusting the mean photon number accordingly.

The single pair per pulse generation probability and the overall probability of success is shown in Figure~\ref{fig:16SourceBar}. The relative increase in SNR by moving from binary to PNR detectors allows each individual source to operate at a higher mean photon number whilst maintaining the same SNR. Coupled with the gains made through multiplexing, this result in a significant increase in the overall success of the system.

The effect of detector efficiency on achievable SNR can be observed in Figure~\ref{fig:16_SourceSNR_nd}. We fix the mean photon number to the value that yields a SNR of 100 at $\eta_{d} = 0.7$ for a 16-way multiplexed source using PNR detection, and calculate the dependence of the SNR on the detector efficiency for the three detector types. As expected, the SNR of the system increases with increasing detector efficiency. The rapid increase in SNR shown by the PNR detectors is again due to the ability to discriminate multi-photon heralding detection events and remove these from the noise; whereas the lack of this information from binary detectors yields a flatter curve. We note that for low mean photon numbers and lumped detection efficiencies up to 80\% the performance of the 8-bin pseudo-PNR detector is almost indistinguishable from that of a PNR detector.

For a PNR detector, the complete dependence of SNR on detection efficiency and mean photon number is shown in Figure~\ref{fig:16_SourceSNR_ColorMap}. Unsurprisingly the best case is when the detector has unit efficiency; this results in an infinite SNR as one can discriminate perfectly between single and multi-photon components. Poor detection efficiency can to some extent be mitigated by operating at low mean photon numbers to reduce noise, but at the expense of overall generation rate. Nevertheless, the generation rate can be recovered through multiplexing as we will demonstrate.

\begin{figure} 
	\subfigure[]{
		\label{fig:Bin_Bar}
		\includegraphics[width=0.22\textwidth]{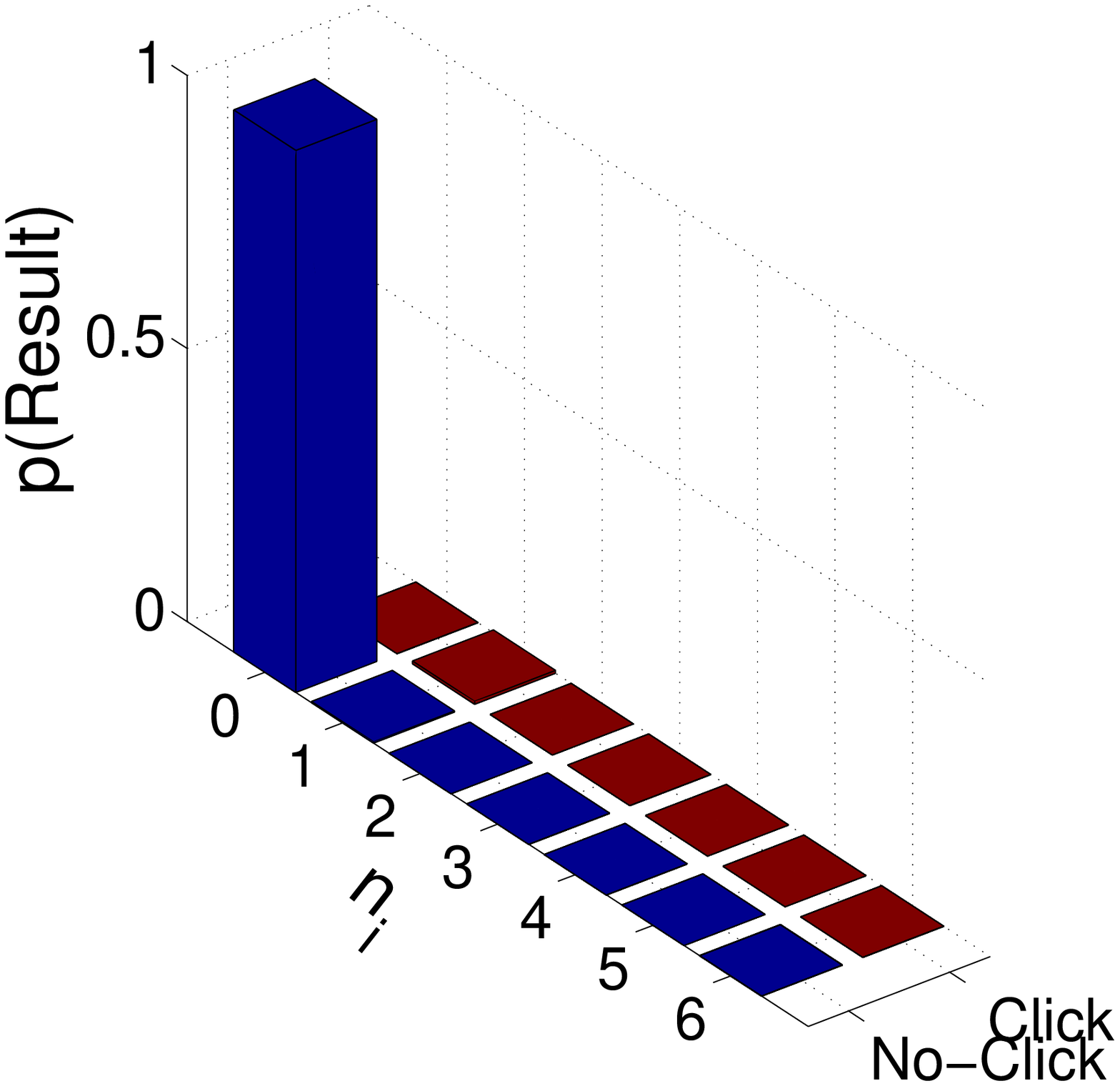}
		}
	\hspace*{\fill}
		\subfigure[]{
		\label{fig:Bin_Bar_16}
		\includegraphics[width=0.22\textwidth]{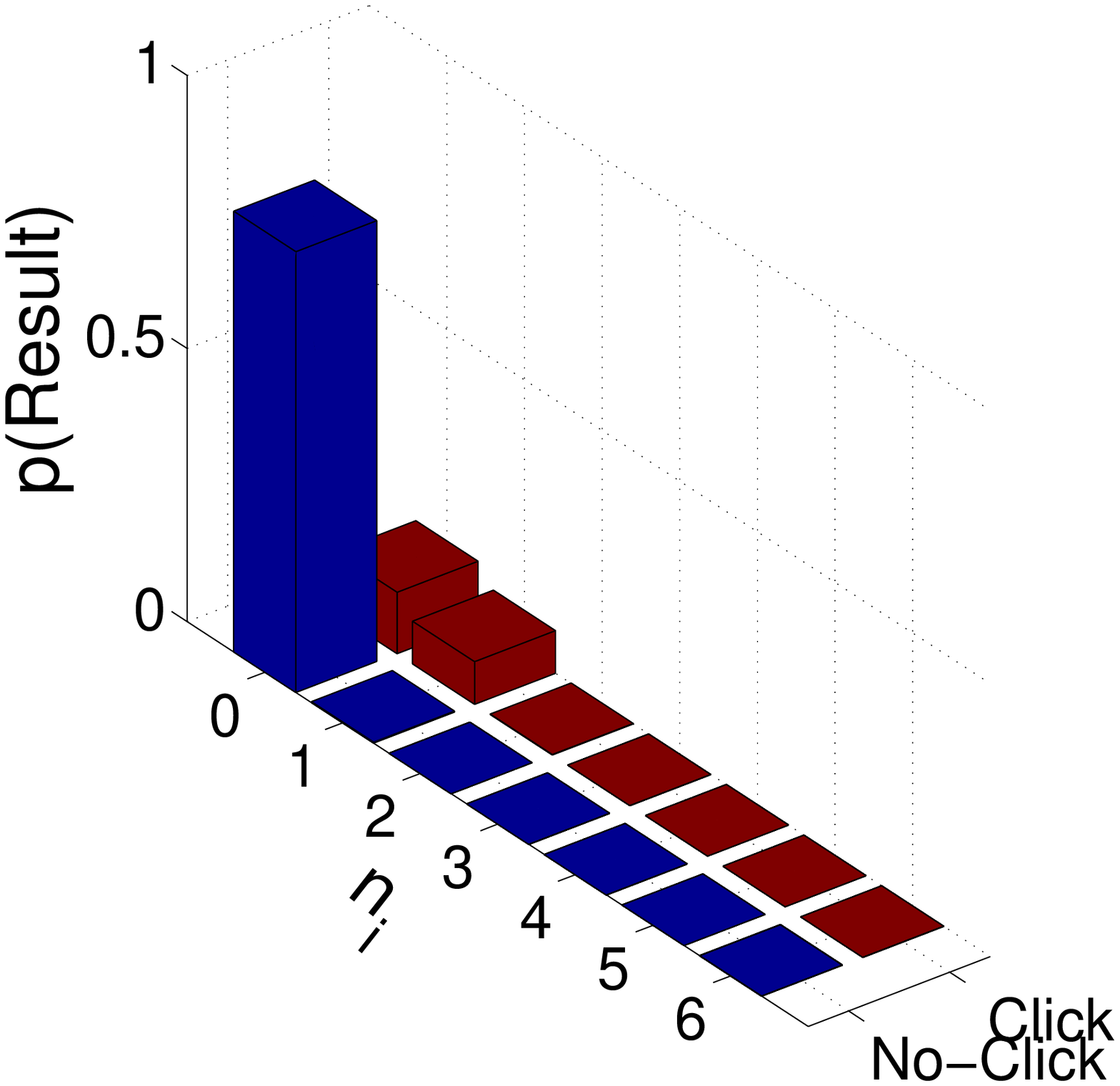}
		}
	
	\subfigure[]{
		\label{fig:PNR_Bar}
		\includegraphics[width=0.22\textwidth]{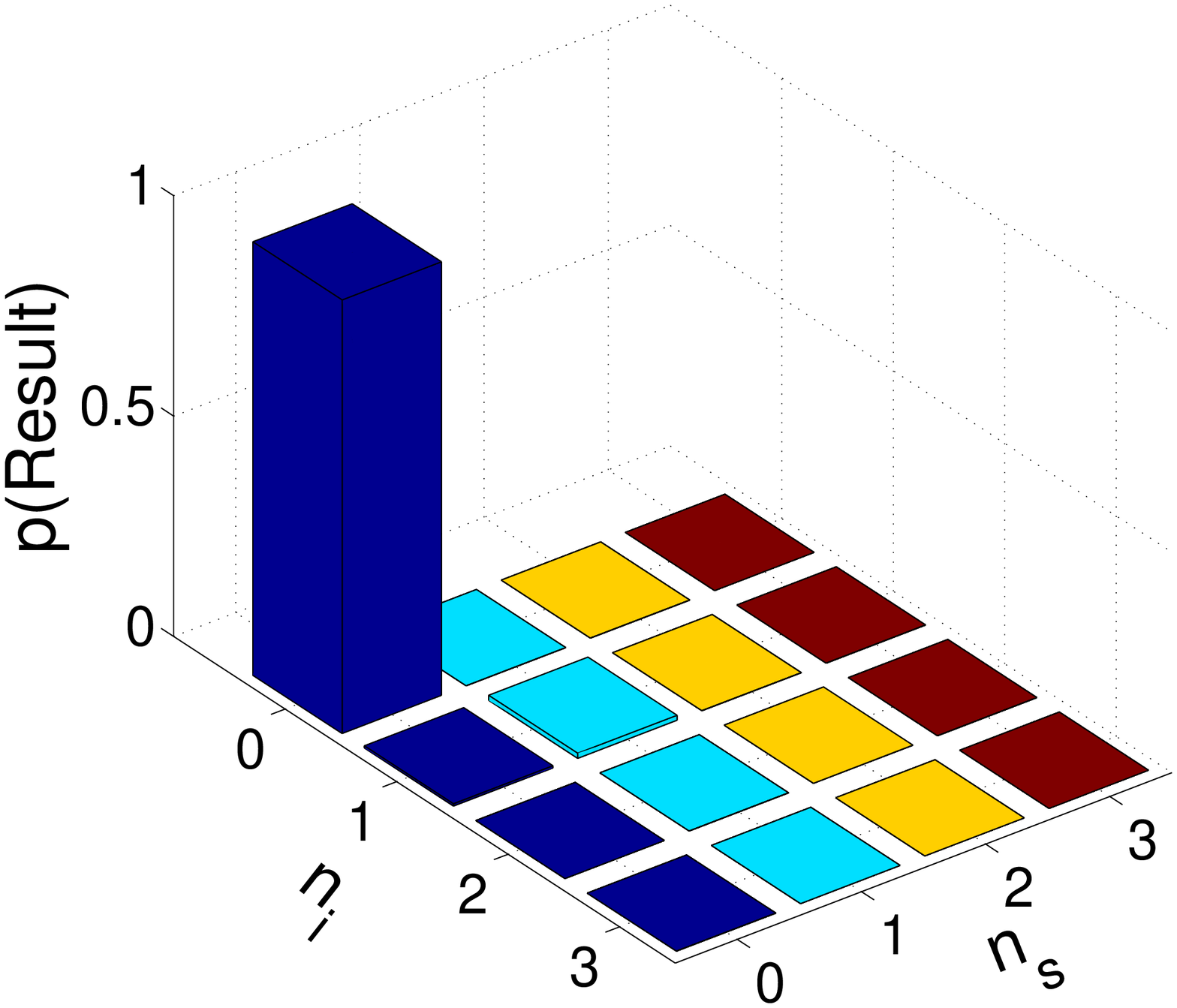}
		}
	\hspace*{\fill}
	\subfigure[]{
		\label{fig:PNR_Bar_16}
		\includegraphics[width=0.22\textwidth]{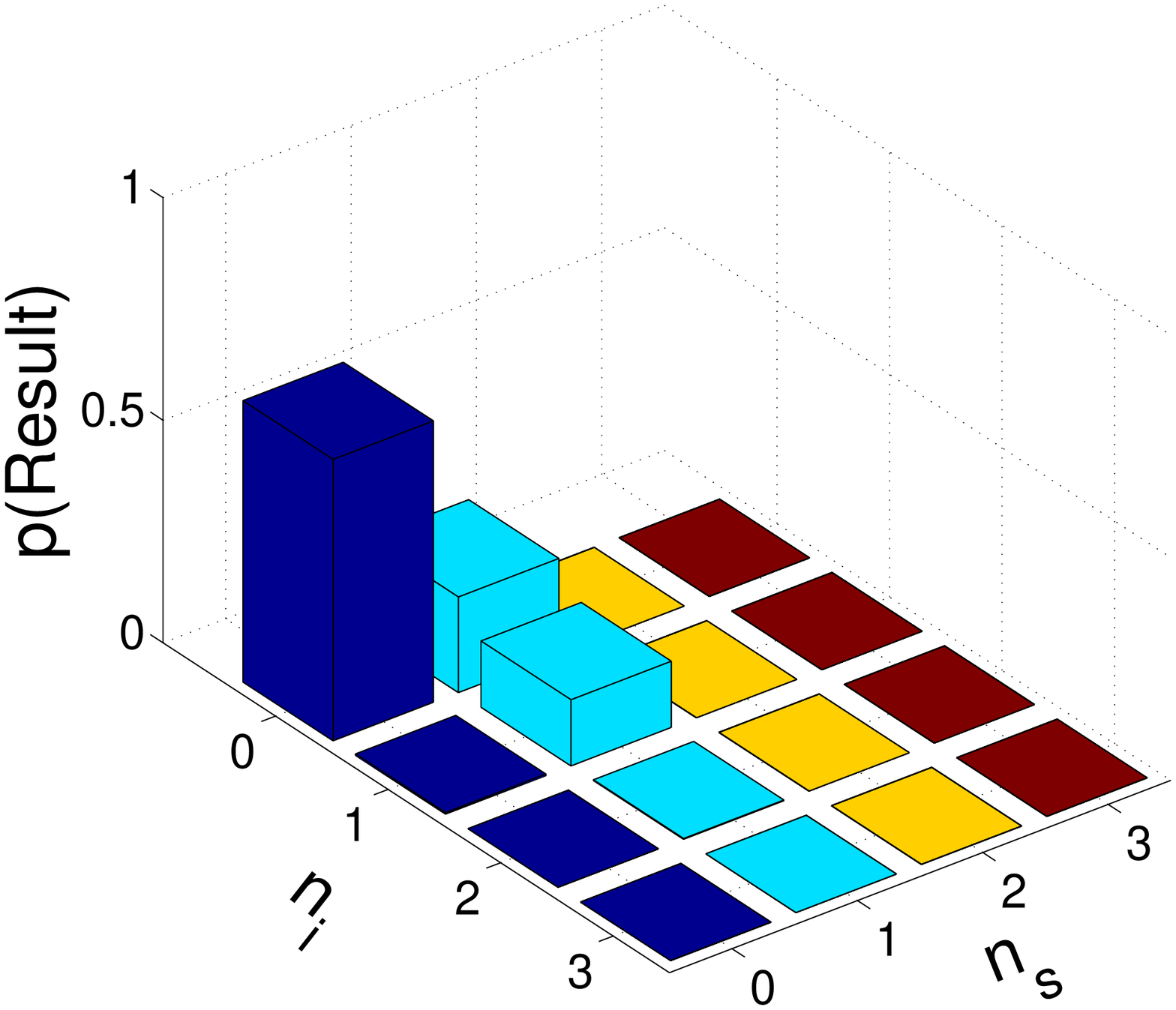}
		}
	\caption{Probability of delivering $n$ photons given a particular heralding event at $\bar{n}$ corresponding to a SNR of 100. Horizontal axes denote the resultant idler state and heralding detection result, vertical axis denotes the overall probability. Efficiencies of $\eta_{d} = 70~\%$, $\eta_{\tau} = 99~\%$ and $\eta_{s} = 80~\%$. a.) Single source binary detector, b.) 16-way multiplexed system with binary detector, c.) Single source PNR detector, d.) 16-way multiplexed system with PNR detector}
\label{fig:BarPlots}
\end{figure}

Figure~\ref{fig:BarPlots} shows the probabilities of delivering a certain number of photons from the source given particular heralding outcomes, at a fixed SNR of 100. For example, the probability of delivering one photon given a successful heralding event is given by Equations \eqref{eq:PNR_pSuccess} and \eqref{eq:Binary_pSuccess} for PNR and binary detectors respectively. To minimise the contributions of higher-order photon-number components when using binary detectors one must operate at low mean photon numbers as seen above. This results in a source in which a single pump pulse will most probably yield nothing at all; only a small fraction of pulses will the detector ``click", as seen for a single binary source in Figure~\ref{fig:Bin_Bar}. By constructing a 16-way multiplexed system with binary detectors the probability of successfully a heralding a single photon is increased compared to the single source as see in Fig.~\ref{fig:Bin_Bar_16}.
In moving to a single PNR source, the probability of success increases slightly over the binary system (Figure~\ref{fig:PNR_Bar}), and the effect of multiplexing 16 sources on the overall probability of success is shown in Figure~\ref{fig:PNR_Bar_16}.

\subsection{PNR detector efficiency}

\begin{figure*}
	\subfigure[]{
		\label{fig:ns_80_nd_30}
		\includegraphics[width=0.7\textwidth]{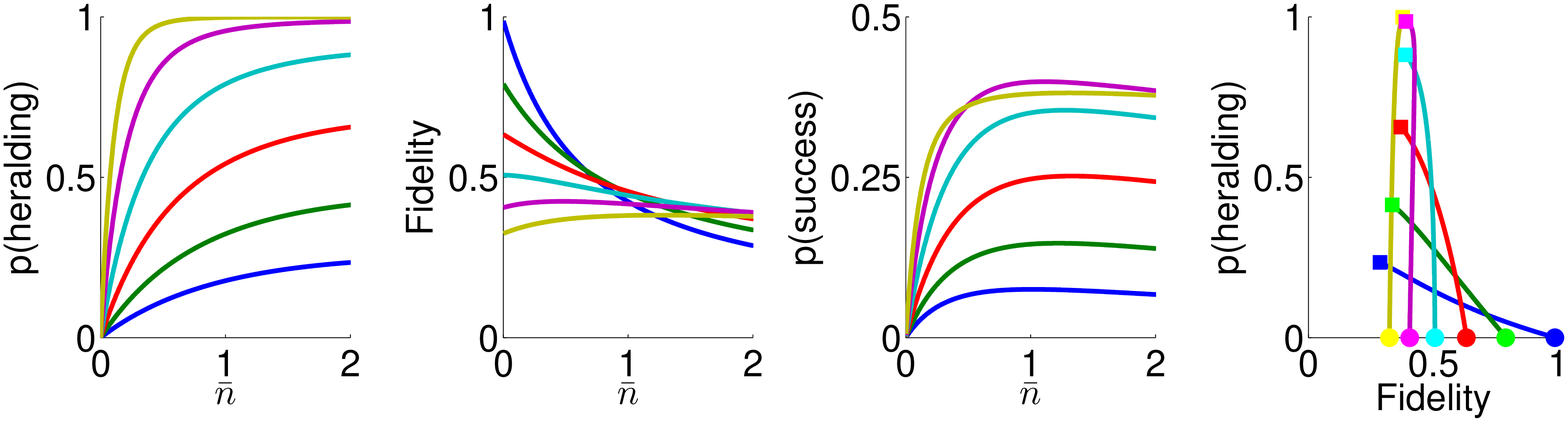}
		}

	\subfigure[]{
		\label{fig:pH_vs_Fidelity_b}
		\includegraphics[width=0.7\textwidth]{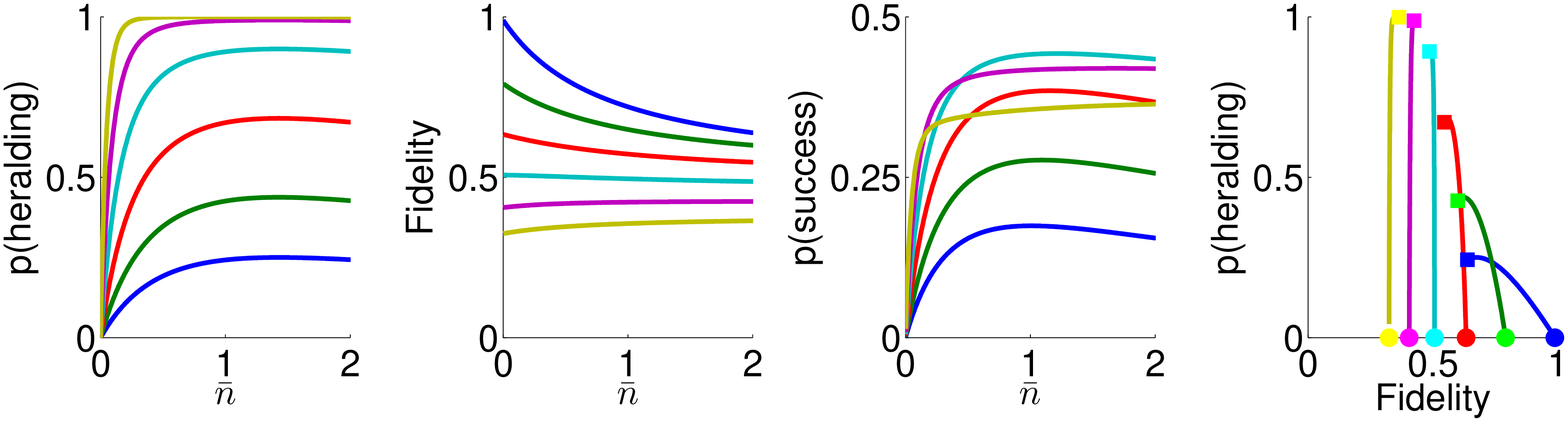}
		}
	
	\subfigure[]{
		\label{fig:pH_vs_Fidelity_c}
		\includegraphics[width=0.7\textwidth]{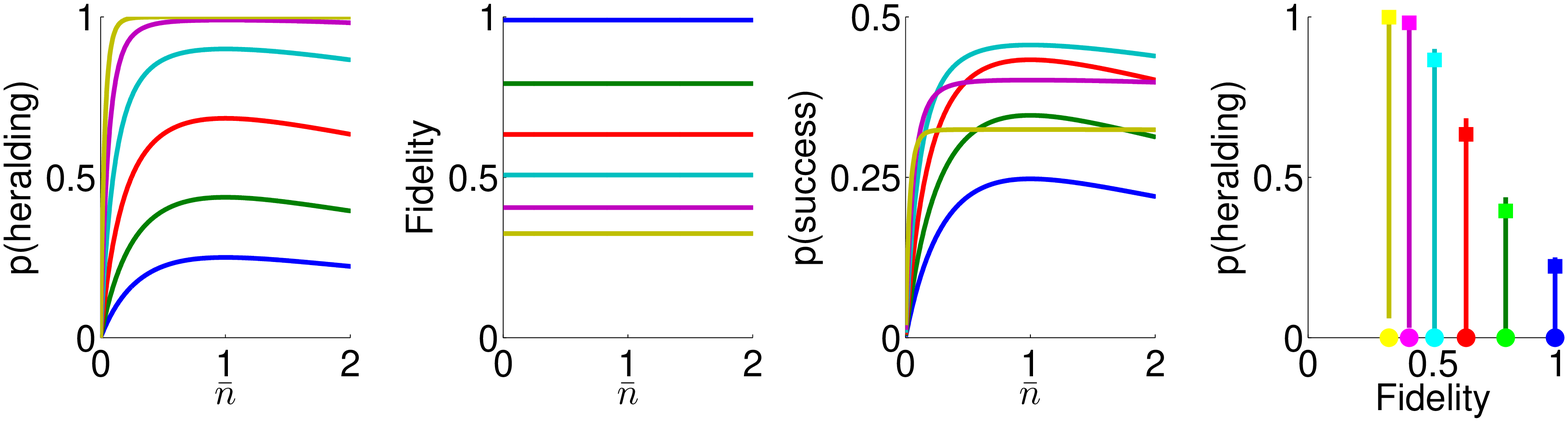}
		}	
	\caption{Multiplexed source performance for varying detector efficiency with switch efficiency $\eta_s = 0.8$. Top row: detector efficiency $\eta_d = 0.3$; middle row: $\eta_d = 0.7$; bottom row: $\eta_d = 1$. First column: probability of successful heralding event, $p(\text{heralding})$, as a function of mean photon number $\bar{n}$; second column: fidelity of output state as a function of $\bar{n}$; third column: probability of successfully delivering a single photon, $p(\text{success}) = p(\text{heralding}) \times \text{fidelity}$, as a function of $\bar{n}$; fourth column: tradeoff between $p(\text{heralding})$ and $\text{fidelity}$ between $\bar{n} = 0$ (circles) and $\bar{n} = 2$ (squares). Colours indicate number of individual sources in each multiplexed system: 1 (blue), 2 (green), 4 (red), 8 (cyan), 16 (purple) and 32 (yellow).}
	\label{fig5}	
\end{figure*}

We use the specific example of PNR detection to investigate the effect of detector efficiency on source performance. Fig~\ref{fig5} shows how the heralding probability, fidelity, and success probability vary as a function of mean photon number for three different detector efficiencies for a fixed switch efficiency. We see straightforwardly that the probability of a successful heralding event increases with the number of individual sources in each multiplexed system, and the initial increase in $p(\text{heralding})$ with $\bar{n}$ becomes more rapid as the detector efficiency improves. The maximum value of $p(\text{heralding})$, which for a perfect detector will always occur at $\bar{n} = 1$ (as the highest probability of single-pair generation for a thermal source occurs at $\bar{n} = 1$, Fig.~\ref{fig:ThermalStats}), moves to higher mean photon numbers as detector efficiency drops and the detectors begin to miss single-pair events and provide spurious ``successful'' heralding signals that correspond to multi-pair inputs. The dependence of the fidelity is more complex. Perfect detectors can always distinguish between single- and multi-pair events, so when $\eta_d = 1$ the fidelity is independent of photon number; however, the switch loss in the idler arm limits the fidelity of the delivered state. Hence in this case the fidelity is constant at the value of the transmission efficiency of the idler arm, which drops as the level of multiplexing increases and more switches are added. When $\eta_d < 1$, as $\bar{n} \rightarrow 0$ the contributions from multi-pair events which would otherwise degrade the fidelity become negligible, and the fidelity assumes the same value as in the case of perfect detection. For a single source, as $\bar{n}$ increases, multi-pair events begin to reduce the fidelity below its $\bar{n}=0$ level as the detector yields more spurious heralding signals. However, for many multiplexed sources, as $\bar{n}$ is increased, the combined effect of detector inefficiency and increased switch loss result in a increase in fidelity over a small range of photon number: a multi-pair event can be incorrectly labelled as a successful event by the heralding detector and then all but one of the photons can be lost in the switch network, resulting in the probabilistic conversion of a multi-pair generation events into a successful outcomes.

The complex interplay between heralding probability and fidelity results in sometimes counter-intuitive behaviour in the success probability; this can be seen in the third column of Fig~\ref{fig5}. For example, an ideal multiplexed source with unit detection efficiency will perform best when operated at a mean photon number of 1 where the probability of generating a single pair assumes its maximum value of 0.25, however we see that the highest performance of a system incorporating imperfect detectors occurs at a mean photon number that does not coincide with the peak probability of generating a single pair from each individual source. Furthermore, we see that for sources with high detection efficiencies with a switch efficiency of less than 100\%, when operating at mean photon numbers above 0.1 the probability of success does not necessarily increase with the number of multiplexed sources. This is a result of the increased loss from the point of generation to the output as extra switch stages are incorporated. The effect of saturation at the total system loss level can be clearly seen where the detector operates at unit efficiency. The limiting factor here is the reduction of fidelity caused by the loss in the heralded photon routing channel which reduces the probability of successfully delivering a single photon from the multiplexed system.

\subsection{Switch efficiency}

\begin{figure*}
	\subfigure[]{
		\label{fig:PNR_pS_nd30}
		\includegraphics[width=0.7\textwidth]{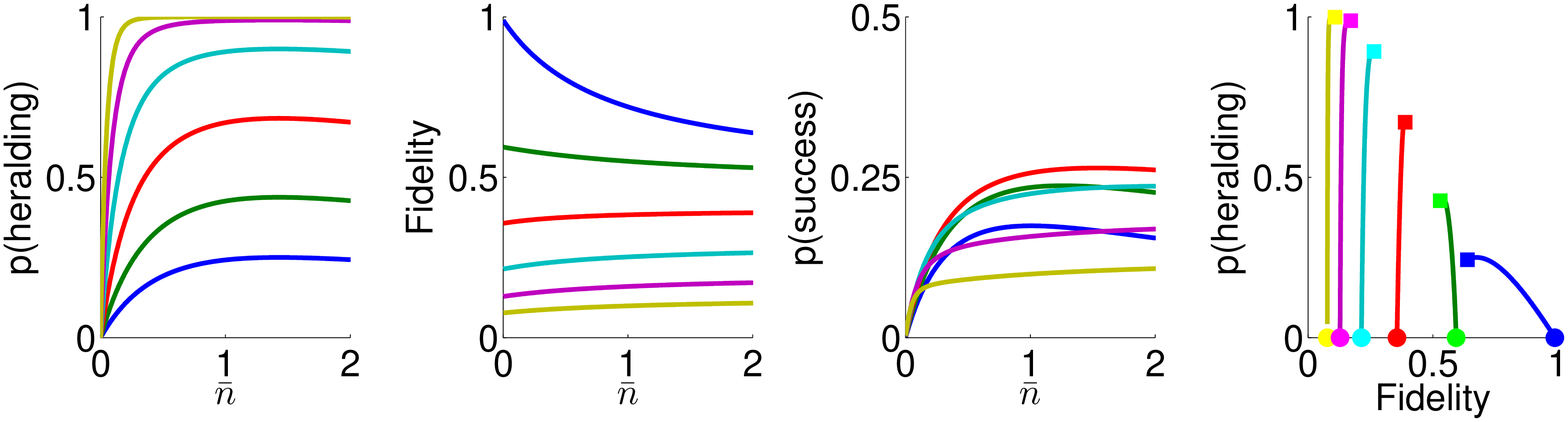}
		}

	\subfigure[]{
		\label{fig:PNR_pS_nd70}
		\includegraphics[width=0.7\textwidth]{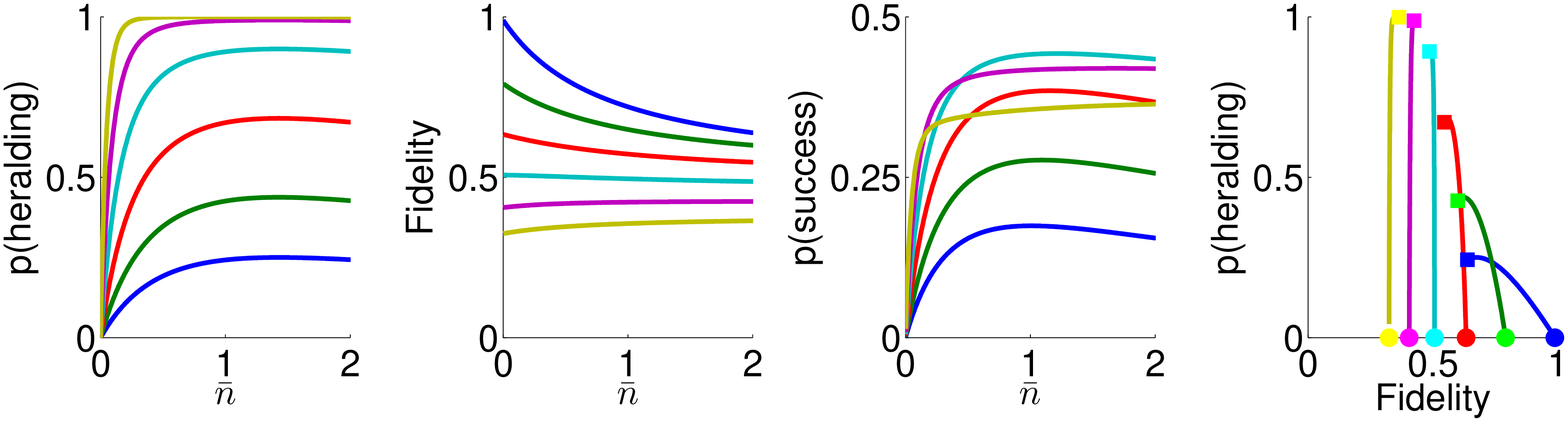}
		}
		
	\subfigure[]{
		\label{fig:PNR_pS_nd100}
		\includegraphics[width=0.7\textwidth]{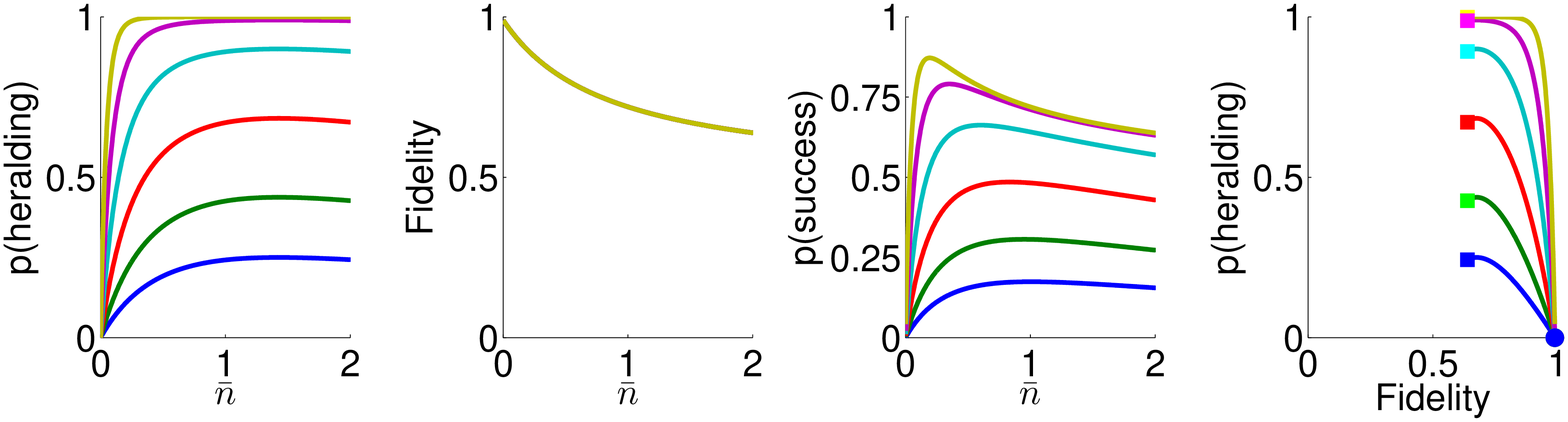}
		}

	\caption{Multiplexed source performance for varying switch efficiency with detector efficiency $\eta_d = 0.7$. Top row: switch efficiency $\eta_s = 0.6$; middle row: $\eta_s = 0.8$; bottom row: $\eta_s = 1$. Column and colour format as in Fig.~\ref{fig5}.}
	\label{fig6}
\end{figure*}

The complete implications of non-unit switch efficiency can be seen in Fig~\ref{fig6}. The heralding probability in the first column does not vary with swith efficiency as the detector efficiency remains fixed at $\eta_d = 0.7$. Similarly the fidelity for a single source does not change (as no switches are used), and for perfect switch transmission the fidelity is independent of the number of stages in the switch network. Again the fidelity at $\bar{n} = 0$ is clamped at the total transmission of the switch network, and similar loss effects can transform multi-pair events into single photon outcomes in the presence of high levels of loss.

The results of this are seen in the success probability in the third column. For high switch loss, adding additional multiplexing stages is only beneficial up to 4 sources; following that the concatenated loss in the switch network limits the fidelity. On the other hand, for high switch transmission, many multiplexing stages yield increases in the success probability. Regardless of any other parameters, reducing the switch loss will always increase $p(\text{success})$; hence it is critical to high-performance operation.

\subsection{\label{sec:OptimisationMulti} Optimisation of Multiplexed Sources}

In order to fully exploit the potential of source multiplexing to achieve near deterministic operation, the efficiency of all the lossy components must be as near to unity as possible. In the ideal case for a PNR detector with unit detection efficiency, the SNR is infinite as the system can always distinguish between single photon and multi-photon components. This corresponds to the case considered in Ref.~\cite{Christ2012Limits_on_the} in which it was shown that deterministic operation (defined as a single photon heralding probability per pulse $> 0.99$) is achievable utilising 17 sources switched together. In this case it is always beneficial to multiplex as many sources as possible together and pump as hard as possible up to the thermal distribution limit of $p_{th}(1) = 0.25$. The final columns in Figs~\ref{fig5} and \ref{fig6} allows a comparison to be made between our results and those of Christ and Silberhorn; it can be seen that by including the effects of loss and inefficiency, source performance is degraded overall but still consistent with their previous work.

Figures~\ref{fig5} and \ref{fig6} deliver the counter-intuitive message that it is not always beneficial to multiplex more sources or operate at the mean photon number per pulse yielding the highest single pair probability. Given realistic components, the construction of an optimised multiplexed system requires the mean photon number to be set in accordance with the loss of the optical components. Nevertheless, if an SNR of 100 is desired, for achievable lumped detector efficiencies the mean photon number is constrained to be low (the regime in which most current sources operate); in this case it can be seen that it is nearly always beneficial to multiplex as many sources as possible.

\begin{figure}
	\includegraphics[width=0.45\textwidth]{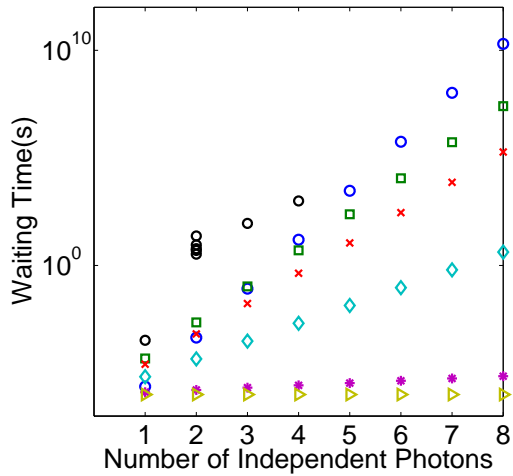}
	\caption{Comparison of waiting times to deliver several independent heralded single photons for different systems at SNR = 100 and 70\% detection efficiency unless otherwise stated. Single sources with binary detection pumped at 80 MHz (blue). Multiplexed sources running at 1\,MHz: 8-way multiplexing with binary detection (green), 4-way multiplexing with pseudo-PNR detection (red), 16-way multiplexing with PNR detection (cyan), 16-way multiplexing with PNR detection for a possible high-performance device with efficiencies $\eta_{d} = 98\%$, $\eta_{\tau} = 99\%$ and $\eta_{s} = 95\%$ (magenta) and deterministic system with lossless routing and switching (yellow). Black circles indicate approximate experimentally measured waiting times for N independent photons for non-multiplexed sources in Ref.~\cite{Mosley2008-Heralded-Generation-of-Ultrafast,Smith20008Heralded_generation_of_two,Lu2007Experimental_Entanglemen_of_Six,Kaltenbaek2006Experimental_Interference,Sun2006Observation_of_the,Pan2012Multiphoton}}
	\label{fig:Wait_Time_Final}
\end{figure}

For realistic components, we have shown that combining individual sources into a multiplexed system can yield significant enhancements in the per-pulse probability of delivering a heralded single photon. We can now calculate the waiting time to deliver $N_{p}$ single photons from $N_p$ independent multiplexed sources, Equation \ref{eq:wait}, and compare it with the equivalent waiting time required for $N_p$ independent single sources. This is shown in Fig.~\ref{fig:Wait_Time_Final}, in which all sources have the same SNR of 100. We see that the waiting time to deliver a small number of heralded single photons can be shorter for non-multiplexed sources; this is because we have allowed these to operate at a pump repetition rate of 80\,MHz, whereas the multiplexed systems are pumped at 1\,MHz to ensure that there is sufficient time for the switches to be set. However, for the delivery of larger numbers of independent heralded photons the multiplexed systems show a drastic reduction in waiting time. The average time to produce 8 heralded single photons simultaneously is reduced from approximately $300$~years for 8 individual sources to a few minutes for 8 realistic multiplexed sources. Finally, we show the performance of a realistic future device, demonstrating the capability of multiplexing to create a near-deterministic source of single photons using only a modest resources in conjunction with achievable PNR detectors.

\section{\label{Conclu} Conclusions}
The work presented here provides a robust manner in which a multiplexed system of single photon sources can be optimised given the inevitable loss of its constituent components. This will be crucial in order achieve the many benefits multiplexing can yield. We have shown the effects of optical loss on the heralded idler state from a single photon source and how this can limit the overall performance of any future multiplexed systems. Nevertheless, we have demonstrated that, even in the presence of imperfect switches and detectors based on current technology, source multiplexing provides a route to huge increases in the performance of heralded single photon sources. Combined with the inevitable improvements in both switches and detectors in the coming years, source multiplexing is a promising candidate for supplying high-quality single photons for future quantum technologies.

\section{Acknowledgements}
We acknowledge support from the UK Engineering and Physical Sciences Research Council under grant EP/K022407/1.

During the preparation of this manuscript we became aware of related work by Adam et al \cite{Adam2014Optimization-of-periodic-single-photon}.

\bibliography{Bibliography}

\end{document}